\documentclass[12pt]{article}
\usepackage{graphicx}

\begin{document}
 
\title{EFFECTIVE CHIRAL THEORY OF LARGE $N_C$ $QCD$ OF MESONS
}
 
\author{B.A.Li\footnote{invited plenary talk at Int.Conf. of Flavor Physics 
2001, Zhang Jia-Jie, Hunan, China, 5/31-6/7}\\
Department of Physics, University of Kentucky,
Lexington,\\ KY 40506, USA\\E-mail: li@pa.uky.edu}

\maketitle
\begin{abstract}
An effective chiral theory of pseudoscalars, vectors,
and axial-vectors
is presented. In this theory mesons are coupled to quarks. At quark level mesons
have no
kinetic terms which are generated by quark loops.
The theory has both explicit chiral symmetry(in the limit, $m_q\rightarrow 0$)
and dynamical chiral symmetry breaking. Large $N_C$ expansion is a natural result.
Tree
diagrams are at leading order and meson loops are at higher orders. In this theory
octet pseudoscalars are Goldstone bosons. Masses and decay widths of the mesons
agree with
data. VMD, Wess-Zumino-Witten anomaly, and OZI rule are obtained. Besides
three current
quark masses there are other two parameters: a universal coupling constant
and a cutoff.
At low energy the theory goes back to ChPT. The phenomenology of the theory
is successful.
\end{abstract}
\newpage 
\section{Introduction}
 
QCD is the theory of strong interactions. The nonperturbative QCD remains unsolved.
Before QCD created there were already a lot of successful studies on meson physics:
Vector Meson Dominance(VMD), Current algebra, PCAC,
Sum rules, Goldstone theorem and Gell-Mann, Oaks, and Renner formula of pion
mass, ABJ and WZW anomaly etc..
 
In order to solve nonperturbative QCD many models and effective
theories have been proposed, for instance,
Quark models from which we learned that there is a constituent quark
mass, Nonlinear $\sigma$ model, NJL model with four fermion interactions,
QCD sum rule, Instanton induced model, Chiral perturbation theory(ChPT),
many more.
 
Mesons are bound states of quarks and gluons. Bound state is a nonperturbative
problem. It is heuristic to revisit $QED$. Hydrogen atom is a bound state
of proton and
electron. We divide $QED$ into two parts: nonperturbative and
perturbative $QED$.
\begin{equation}
{\cal L}_{QED}={\cal L}_{nonperturbative QED}+{\cal L}_{perturbative QED}.
\end{equation}
The interaction in ${\cal L}_{nonperturbative QED}$ is mainly coulomb interaction,
while in perturbative part is mainly transverse photon. Boglubov has proposed a
rigorous method to do the separation of $QED$.
We use ${\cal L}_{nonperturbative QED}$ to get bound state solution,
while use the ${\cal L}_{perturbative QED}$ to do radiative correction.
 
Of course, $QCD$ is different
from $QED$.Gluon fields are nonabelian fields.
However, $QCD$ is still divided into nonperturbative and perturbative
parts.
\begin{equation}
{\cal L}_{QCD}={\cal L}_{nonperturbative QCD}+{\cal L}_{perturbative QCD}.
\end{equation}
In ${\cal L}_{nonperturbative QCD}$ soft gluons and instantons play dominant roles,
while it is well known that ${\cal L}_{perturbative QCD}$ is dominant by hard gluons.
Unlike $QED$ we do not know the explicit expression of
${\cal L}_{nonperturbative QCD}$. There are different approaches to study
${\cal L}_{nonperturbative QCD}$. Effective theory is one of them.
 
In $QCD$ mesons are bound states of quarks and gluons. Meson physics at low
energy is nonperturbative. We have proposed a chiral theory of mesons[1], which is an
effective theory of nonperturbative part of $QCD$,
to treat meson problems. The perturbative part of $QCD$ is used to calculate the
radiative correction by hard gluons.
 
In this talk we focus on the effective theory of mesons.
 
\section{\bf Effective large $N_C$ QCD of mesons}
The strategy of constructing the effective theory of nonperturbative $QCD$ of mesons
is following.
\begin{enumerate}
\item Mesons are coupled to quarks.\\
In $QCD$ meson is a pole of four point Green function of quarks. A Green function of
many quark pairs can be separated into product of corresponding four point
Green functions
of quark pairs and kernel. Mesons are poles of the four point Green functions.
At the poles
mesons are coupled to quarks(kernel). Based on this picture we try to construct an
effective Lagrangian to calculate this part of the Green function.
\item From $SU(3)_L\times SU(3)_R$ current algebra we have learned that vector
and axial-vector mesons are coupled to vector and axial-vector currents of quarks
respectively.
\item We use the scheme of nonlinear $\sigma$ model to introduce pseudoscalars to
the theory.
\item The theory has explicit chiral symmetry in the limit of \(m_q\rightarrow 0\),
where $m_q$ is the current quark mass.
Therefore, in the chiral limit pseudoscalars are massless and vector and axial-vector
mesons have the same masses.
\item $QCD$ has dynamical chiral symmetry breaking. The dynamical chiral
symmetry breaking should be introduced to the theory.
\end{enumerate}
 
Based on the inputs mentioned above, for two flavors
the effective Lagrangian of pseudoscalar, vector
, axial-vector mesons has been constructed as[1]
\begin{eqnarray}
{\cal L}=\bar{\psi}(x)(i\gamma\cdot\partial+\gamma\cdot v
+\gamma\cdot a\gamma_{5}
-mu(x))\psi(x)-\bar{\psi(x)}M\psi(x)\nonumber \\
+{1\over 2}m^{2}_{0}(\rho^{\mu}_{i}\rho_{\mu i}+
\omega^{\mu}\omega_{\mu}+a^{\mu}_{i}a_{\mu i}+f^{\mu}f_{\mu})
\end{eqnarray}
where \(a_{\mu}=\tau_{i}a^{i}_{\mu}+f_{\mu}\), \(v_{\mu}=\tau_{i}
\rho^{i}_{\mu}+\omega_{\mu}\),
and \(u=exp\{i\gamma_{5}(\tau_{i}\pi_{i}+
\eta)\}\), and M is the current quark mass matrix.
Since mesons are bound states solutions of $QCD$ they
are not independent degrees of freedom. Therefore,
there are no kinetic terms for meson fields. The kinetic terms
of meson fields are generated from quark loops.
A similar Lagrangian can be constructed for three flavors. According to $QED$ and the
Standard Model photon, W and Z fields can be incorporated into this Lagrangian.
 
This Lagrangian has following features
\begin{enumerate}
\item
Obviously, in the limit \(m_q\rightarrow 0\) this Lagrangian has chiral symmetry.
\item
In Eq.(3) there is a parameter m, the constituent quark mass, which is
related to quark condensate
\begin{equation}
<\bar{\psi}\psi>={i\over(2\pi)^{4}}
Tr\int d^{4}p\frac{\gamma\cdot p-m\hat{u}}{p^{2}
-m^{2}}
=-\frac{3m^{3}N_{C}}{4\pi^{2}}
\{{\Lambda^{2}\over m^{2}}-log({\Lambda^{2}\over m^{2}}+1)\},
\end{equation}
where $\Lambda$ is the cutoff of this effective theory.
a cutoff is necessary for an effective theory. It should be the match point between
nonperturbative and perturbative QCD. The theory has dynamical chiral symmetry breaking.
\item
The effective Lagrangian of mesons is obtained by integrate out quark fields in Eq.(3)
or by calculating quark loop diagrams. Using the method of path integral,
the effective Lagrangian of mesons
is obtained.
 
\begin{eqnarray}
{\cal L}^{M}_{E}={\cal L}_{RE}+{\cal L}_{IM},\nonumber \\
{\cal L}_{RE}={1\over 2}log det({\cal D}^{\dag}{\cal D}),
\;\;\;{\cal L}_{IM}={1\over 2}
log det({\cal D}/{\cal D}^{\dag})
\end{eqnarray}
where
\begin{equation}
{\cal D}=\gamma\cdot\partial-i\gamma\cdot v-i\gamma\cdot a\gamma_{5}+mu.
\end{equation}
\begin{equation}
{\cal D}^{\dag}=-\gamma\cdot \partial+i\gamma\cdot v
-i\gamma\cdot a\gamma_{5}+m\hat{u},\;\;\;
\hat{u}=exp(-i)\gamma_{5}(\tau_{i}\pi_{i}+\eta).
\end{equation}
In ${\cal L}_{RE}$ there are even numbers of $\gamma_5$, while in ${\cal L}
_{IM}$ the number of $\gamma_5$ is odd. Therefore, ${\cal L}_{IM}$ is the
anomalous action.
 
The kinetic terms of the vector, axial-vector, and pseudoscalar fields are
generated by quark loops and the physical meson fields are defined
\begin{eqnarray}
\pi\rightarrow {2\over f_{\pi}}\pi,\;\;\;\eta\rightarrow
{2\over f_{\eta}}\eta, \;\;\;
\rho\rightarrow {1\over g}\rho,\;\;\;
\omega\rightarrow {1\over g}\omega,
\end{eqnarray}
$f_\pi$ is defined by normalizing the kinetic terms of pseudoscalar fields and the
universal coupling constant,g, is to normalizing the kinetic terms of vector fields.
\begin{equation}
f^2_\pi=F^{2}(1-{2c\over g}),\;\;\;
c=\frac{f^{2}_{\pi}}{2gm^{2}_{\rho}},\;\;\;
g^2={1\over 6}{F^{2}\over m^{2}},
\end{equation}
where
\[{F^2\over16}=\frac{N_C}{(4\pi)^2}m^2\int d^4 p\frac{1}{(p^2+m^2)^2}.\]
 
There are mixing between axial-vectors and corresponding pseudoscalars. The
substitution
\begin{equation}
a^{i}_{\mu}\rightarrow {1\over g}(1-{1\over 2\pi^{2}g^{2}})
^{-{1\over 2}}a^{i}_{\mu}
-{c\over g}\partial_{\mu}\pi^{i}.
\end{equation}
is used to erase the mixing $a^i_\mu\partial_\mu\pi^i$.
 
To the fourth order in covariant derivatives in Minkowsky space the
lagrangian takes the following form
\begin{eqnarray}
\lefteqn{{\cal L}_{RE}=\frac{N_{c}}{(4\pi)^{2}}m^{2}{D\over 4}\Gamma
(2-{D\over 2})TrD_{\mu}UD^{\mu}U^{\dag}}\nonumber \\
 & &-{1\over 3}\frac{N_{c}}{(4\pi
)^{2}}{D\over 4}\Gamma(2-{D\over 2})\{2\omega_{\mu\nu}\omega^{\mu\nu}
+Tr\rho_{\mu\nu}\rho^{\mu\nu}+2f_{\mu\nu}f^{\mu\nu}+
Tra_{\mu\nu}a^{\mu\nu}\}\nonumber \\
 & &+{i\over 2}\frac{N_{c}}{(4\pi)^{2}}Tr\{D_{\mu}UD_{\nu}U^{\dag}+
D_{\mu}U^{\dag}D_{\nu}U\}\rho^{\nu\mu}\nonumber \\
 & &+{i\over 2}\frac{N_{c}}{(4\pi)^{2}}Tr\{D_{\mu}U^{\dag}D_{\nu}U-
 D_{\mu}
UD_{\nu}U^{\dag}\}a^{\nu\mu} \nonumber \\
 & &+\frac{N_{c}}{6(4\pi)^{2}}TrD_{\mu}D_{\nu}UD^{\mu}D^{\nu}U^{\dag}
\nonumber \\
 & &-\frac{N_{c}}{12(4\pi)^{2}}Tr\{
D_{\mu}UD^{\mu}U^{\dag}D_{\nu}UD^{\nu}U^{\dag}
+D_{\mu}U^{\dag}D^{\mu}UD_{\nu}U^{\dag}D^{\nu}U
-D_{\mu}UD_{\nu}U^{\dag}D^{\mu}UD^{\nu}U^{\dag}\} \nonumber \\
 & &+{1\over 2}m^{2}_{0}(\omega_{\mu}\omega_{\mu}+\rho^{i}_{\mu}
\rho^{i\mu}+a^{i}_{\mu}a^{i\mu}+f_{\mu}f^{\mu}),
\end{eqnarray}
where
\begin{eqnarray*}
D_{\mu}U=\partial_{\mu}U-i[\rho_{\mu}, U]+i\{a_{\mu}, U\},\\
D_{\mu}U^{\dag}=\partial_{\mu}U^{\dag}-i[\rho_{\mu}, U^{\dag}]-
i\{a_{\mu}, U^{\dag}\},\\
\omega_{\mu\nu}=\partial_{\mu}\omega_{\nu}-\partial_{\nu}\omega_{\mu},
\\
f_{\mu\nu}=\partial_{\mu}f_{\nu}-\partial_{\nu}f_{\mu},\\
\rho_{\mu\nu}=\partial_{\mu}\rho_{\nu}-\partial_{\nu}\rho_{\mu}
-i[\rho_{\mu}, \rho_{\nu}]-i[a_{\mu}, a_{\nu}],\\
a_{\mu\nu}=\partial_{\mu}a_{\nu}-\partial_{\nu}a_{\mu}
-i[a_{\mu}, \rho_{\nu}]-i[\rho_{\mu}, a_{\nu}],\\
D_{\nu}D_{\mu}U=\partial_{\nu}(D_{\mu}U)-i[\rho_{\nu}, D_{\mu}U]
+i\{a_{\nu}, D_{\mu}U\},\\
D_{\nu}D_{\mu}U^{\dag}=\partial_{\nu}(D_{\mu}U^{\dag})
-i[\rho_{\nu}, D_{\mu}U^{\dag}]
-i\{a_{\nu}, D_{\mu}U^{\dag}\}.
\end{eqnarray*}
All the vertices of mesons are derived from Eq.(11).
\item
Besides three current quark masses,
there are other two parameters in this effective theory: cutoff $\Lambda$ and g.
Input $f_\pi$, the decay rate of $\rho\rightarrow ee^+$,
$\Lambda$ and g are determined, $\Lambda\sim 1.8GeV$ and \(g=0.395\).
three current quark masses are determined by input $m^2_\pi$,
$m^2_{K^{\pm}}$.
\item
$N_C$ comes from the quark loop. We have
\[f^2_\pi\sim O(N_C),
\;\;\; g\sim O(\sqrt{N_C}).\]
All the meson fields
are order of $O(\sqrt{N_C})$ and all the meson vertices are $O(N_C)$.
Therefore, the diagrams at tree level are $O(N_C)$ and loop diagrams are
at higher order in $N_C$ expansion. t'Hooft and Witten have proposed using
$N_C$ expansion to study $QCD$. In this theory
$N_C$ expansion is a natural result. $N_C$ expansion is used to do
physical calculations. The cutoff is about 1.8GeV. Therefore, only low
lying mesons contribute to the loop diagrams. Namely, the loop diagrams
are calculable. This is the reason why this effective theory is named as
effective chiral large $N_C$ $QCD$ of mesons.
\end{enumerate}
Now we can use the vertices derived from Eq.(11) to study meson physics.
\section{Symmetry breaking and Masses of mesons}
In field theory meson masses are always associated with symmetry breaking.
$\pi$, $\rho$, and $a_1$ are all made of u and d quarks\\
{\bf Why pion is very light?}\\
{\bf Why $\rho$ is much heavier than pion?}\\
{\bf Why $a_1$ is heavier than $\rho$?}\\
How to understand Weinberg's second sum role \(m^2_a=2m^2_\rho\)?\\
\begin{enumerate}
\item To the leading order in quark mass expansion,
the masses of the octet pseudoscalar mesons are derived
\begin{eqnarray}
m^{2}_{\pi}=-{2\over f^{2}_{\pi}}(m_{u}+m_{d})<0|\bar{\psi}\psi|0>,
\nonumber \\
m^{2}_{K^{+}}=-{2\over f^{2}_{\pi}}(m_{u}+m_{s})<0|\bar{\psi}\psi|0>,
\nonumber \\
m^{2}_{K^{0}}=-{2\over f^{2}_{\pi}}(m_{d}+m_{s})<0|\bar{\psi}\psi|0>,
\nonumber \\
m^{2}_{\eta}=-{2\over 3f^{2}_{\pi}}(m_{u}+m_{d}+4m_{s})
<0|\bar{\psi}\psi|0>.
\end{eqnarray}
{\bf pion mass originates in explicit chiral symmetry breaking by current
quark masses}
 
On the other hand, it is interesting to point out that there are two
diagrams from the two vertices
\begin{equation}
-im\bar{\psi}\tau^i\gamma_5\psi\pi^i-{1\over2}m\bar{\psi}\psi\pi^2
\end{equation}
contribute to pion mass. The later is a tadpole. There is destructive
interference between these two diagrams. The cancellation leads to the
mass formula above. In the limit $m_q\rightarrow 0$, \(m_\pi=0\).
Goldstone theorem is satisfied.
\item {\bf $m_\rho$}\\
In this theory we have
\begin{equation}
m^{2}_{\rho}=m^{2}_{\omega}={1\over g^{2}}m^{2}_{0}.
\end{equation}
From Eq.(3) we can see that current algebra is satisfied by this theory.
The KSFR sum rule
\begin{equation}
g_{\rho}={1\over 2}f_{\rho\pi\pi}f^{2}_{\pi}
\end{equation}
can be taken as the equation to determine
$m_{\rho}$. we have
\[f_{\rho\pi\pi}={2\over g},\;\;\;
g_{\rho}={1\over 2}gm^{2}_{\rho}.\]
Substituting them into the KSFR sum rule, we obtain
\begin{equation}
m^{2}_{\rho}=2{f^{2}_{\pi}\over g^{2}}=6m^2.
\end{equation}
Therefore,
{\bf $m_\rho$ is resulted in dynamical chiral symmetry breaking}
for the masses of $K^*$ and $\phi$ strange quark mass corrections
should be taken into account.
\item {\bf $m_{a_1}$}\\
In the original Lagrangian because of chiral symmetry both $\rho$
and $a_1$ have the same mass. $\rho$ is coupled to vector quark current and
$a_1$ couples to axial-vector current of quarks
\[\bar{\psi}\tau^i\gamma_\mu\gamma_5\psi a^i_\mu.\]
When we calculate the vacuum polarization diagram, unlike the vector
coupling, an additional mass term is generated and chiral symmetry is
broken by the axial-vector coupling. We name this symmetry breaking
as {\bf axial-vector symmetry breaking}.
We obtain
\begin{equation}
(1-{1\over 2\pi^2g^2})m^2_a={F^2\over g^2}+m^2_\rho=2m^2_{\rho}.
\end{equation}
The left hand is the result of Weinberg's $2^{nd}$ sum rule. Here we have
a new factor. In deriving the $2^{nd}$ sum rule Weinberg made an assumption
about the high energy behavior of the propagator. The factor on the left hand side
of Eq.(17) is originated in the high energy behavior of the propagator of $a_1$ field.
This formula fits the data better.
\[m_{a_1}=1.2GeV\;\;\;
m_{a_1}=1.09GeV(Weinberg)\;\;\;
m_{a_1}=1.23\pm0.04GeV(data)\]
\end{enumerate}
There are similar formulas for $f_1(1280)$, $K_1$ and $f_1(1510)$
\begin{equation}
(1-{1\over 2\pi^2g^2})m^2_{f1(1280)}={F^2\over g^2}+m^2_\omega.
\end{equation}
\begin{equation}
(1-{1\over 2\pi^2g^2})m^2_{K_1(1400)}={F^2\over g^2}+m^2_{K^*}.
\end{equation}
\begin{equation}
(1-{1\over 2\pi^2g^2})m^2_{f_1(1510)}={F^2\over g^2}+m^2_{\phi}.
\end{equation}
These results agree with data well.
 
Therefore, there are three symmetry breaking in this theory: explicit chiral
symmetry breaking, dynamical chiral symmetry breaking, and axial-vector symmetry
breaking.
 
In the SM intermediate bosons couple to both vector and axial-vector currents of
fermions. There are axial-vector symmetry breaking which lead to
\begin{enumerate}
\item
\[m^2_W={1\over2}g^2 m^2_t,\;\;\;m^2_Z={1\over2}(g^2+g'^2)m^2_t,[9]\]
\item Two charges and one neutral spin-0 particles exist, whose masses are
$\sim 10^{14}$Gev and they are unphysical. Unitarity of the SM is broken at
$10^{14}$ GeV[10].
\end{enumerate}
 
By the way, Weinberg's first sum rule
\begin{equation}
{g^{2}_{\rho}\over m^{2}_{\rho}}-{g^{2}_{a}\over m^{2}_{a}}={1\over
4}f^{2}_{\pi},
\end{equation}
is satisfied analytically.
 
\section{Normal strong Decays}
The vertices are found from the Lagrangian of mesons. The amplitudes
of decays are calculated in the chiral limit, at the tree level, and up
to the fourth order in derivatives.
\subsection{\bf $V\rightarrow PP$ decays}
Taking $\rho\rightarrow \pi\pi$ as an example.
\begin{eqnarray}
{\cal L}_{\rho\pi\pi}=f_{\rho\pi\pi}\epsilon_{ijk}\rho^{\mu}_{i}
\pi_{j}\partial_{\mu}\pi_{k},\nonumber \\
f_{\rho\pi\pi}={2\over g}\{1+\frac{m^{2}_{\rho}}{2\pi^{2}f^{2}_{\pi}}
[(1-{2c\over g})^{2}-4\pi^{2}c^{2}]\}.
\end{eqnarray}
There is a form factor $f_{\rho\pi\pi}$.
\[\Gamma_\rho=143.2MeV,\;\;exp.=150MeV,\]
We obtain
\[\Gamma(K^*\rightarrow K\pi)=44.9MeV,\;\;exp.=(49.8\pm.8)MeV,\]
\[\Gamma(\phi\rightarrow KK)=3.54 MeV;\;\exp.=3.69(1\pm0.028)MeV.\]
 
The form factor $f^2_{\rho\pi\pi}(q^2)$ contributes $34\%$, $46\%$, and
$61\%$
to the three decays respectively.
\subsection{$A\rightarrow VP$ decays}
\begin{eqnarray}
\lefteqn{{\cal L}_{a_{1}\rightarrow\rho\pi}=
\epsilon_{ijk}\{Aa^{\mu}_{i}\rho_{j\mu}\pi_{k}
+Ba^{\mu}_{i}\rho^{\nu}_{j}\partial_{\mu\nu}\pi_{k}\},}\nonumber \\
&&A={2\over f_{\pi}}(1-{1\over 2\pi^{2}g^{2}})^{-{1\over 2}}
\{{F^{2}\over g^{2}}+\frac{m^{2}_{a}}
{2\pi^{2}g^{2}}-[{2c\over g}+{3\over 4\pi^{2}g^{2}}(1-{2c\over g})]
(m^{2}_{a}-m^{2}_{\rho})\}\nonumber \\
&&B=-{2\over f_{\pi}}(1-{1\over 2\pi^{2}g^{2}})^{-{1\over 2}}
{1\over 2\pi^{2}g^{2}}(1-{2c\over g}).
\end{eqnarray}
The width of the decay is calculated to be
326MeV which is comparable with data(about 400MeV).
There are s-wave and d-wave in this
decay
\begin{equation}
{d\over s}=-0.1.
\end{equation}
The experimental value is $-0.11\pm 0.02$.
We also obtain
\[\Gamma(K_1(1400)\rightarrow K^*\pi)=126MeV,\;\;exp.=163.6(1\pm0.14)MeV,\]
\[\Gamma(f_1(1510)\rightarrow K^* K)=22MeV,\;\;exp.=35\pm15MeV,\]
Amplitude A plays dominant role, in which there is cancellation between
the two terms of A. In 60's current algebra has difficulty to get the
decay width of $a_1$ meson. This effective theory provides an explanation.
In this theory in unphysical region the current algebra result is
satisfied. However, because of the cancellation in the amplitude A the
physical region is far away from unphysical.
 
The decay $\eta'\rightarrow\eta\pi\pi$ is calculated too.
\[\Gamma(\eta'\rightarrow\eta\pi^+\pi^-)=85.7keV, exp.=
87.8\pm0.12 keV,\]
\[\Gamma(\eta'\rightarrow\eta\pi^0\pi^0)=48.6keV, exp.=
41.8\pm0.11 keV.\]
\section{\bf Vector meson dominance}
The interactions between photon and mesons are derived as
\begin{equation}
{e\over f_{\rho}}\{-{1\over 2}F^{\mu\nu}(\partial_{\mu}\rho^{0}_
{\nu}-\partial_{\nu}\rho^{0}_{\mu})+A^{\mu}j^{0}_{\mu}\}.
\end{equation}
\begin{eqnarray}
{e\over f_{\omega}}\{-{1\over 2}F^{\mu\nu}(\partial_{\mu}\omega_
{\nu}-\partial_{\nu}\omega_{\mu})+A^{\mu}j^{\omega}_{\mu}\},
\nonumber \\
{e\over f_{\phi}}\{-{1\over 2}F^{\mu\nu}(\partial_{\mu}\phi_
{\nu}-\partial_{\nu}\phi_{\mu})+A^{\mu}j^{\phi}_{\mu}\},
\end{eqnarray}
where
\begin{equation}
{1\over f_{\rho}}={1\over 2}g,\;\;\;
{1\over f_{\omega}}={1\over 6}g,\;\;
{1\over f_{\phi}}=-{1\over 3\sqrt{2}}g
\end{equation}
 
These are the expressions of VMD proposed by Sakurai.
$\rho\rightarrow ee^+$ is used to determine \(g=0.395\).
 
\section{\bf Form factors}
For a long time pion form factor is expressed as a $\rho$ pole. However,
The radius from $\rho$ pole is less than the data by about $10\%$.
Comparing with data, the
$\rho$ pole form factor decreases slower in time like region and faster
in space like region.
 
We have studied the form factors of pion and kaon[3]. The pion
form factor is expressed as
\begin{eqnarray}
|F_\pi (q^2)|^2 & = & f^2_{\rho \pi \pi }(q^2)
\frac{m_\rho ^4+q^2\Gamma _\rho
^2(q^2)}{(q^2-m_\rho ^2)^2+q^2\Gamma _\rho ^2(q^2)}
\end{eqnarray}
The
radius of charged pion is found to be
\begin{equation}
<r^2>_\pi =(0.395+0.057)fm^2 = 0.452fm^2.
\end{equation}
The contribution of the intrinsic form
factor,$f_{\rho\pi\pi}$, is about $13\%$ of the total value. The
experimental data is $(0.439\pm 0.03)fm^2$.
 
The decay rate of
$\tau\rightarrow\pi\pi\nu$ is dominated by pion form factor, which agrees
well with CLEO.
We obtain
$B_{\tau ^{-}\rightarrow
\pi ^0\pi ^{-}\nu _\tau }=22.3\%.$
The experimental data is
$(25.32\pm 0.15)\%$.
 
In the same way, kaon form factors are obtained. The radii of kaons are
calculated
\[<r^2>_{K^{+}}=
0.38fm^2.\;\;\;
<r^2>^{exp}_{K^+}=
=0.34\pm0.05 fm^2.\]
The decay rate of
$\tau^-\rightarrow K^0 K^-\nu$ is dominated by pion form factor too.
We obtain \(B=1.78\times
10^{-3}\).
The data is $(1.59\pm 0.24)\times 10^{-3}$.
 
The form factors of pion and charged kaon agree with data in both 
timelike and spacelike regions below 1.5GeV.
 
The form factors of $K_{l3}$ are dominated by a $K^*$ pole and the
intrinsic form factor $f_{\rho\pi\pi}$. Theory agrees with data.
 
The three form factors of
$\pi\rightarrow e\gamma\nu$ and
$K\rightarrow e\gamma\nu$
are computed. Theory agrees well with data. PCAC is satisfied analytically.
\section{\bf $\pi\pi$ and $\pi K$}
The amplitudes of $\pi-\pi$ scattering are calculated[1]. For
the channel of p-wave and \(I=1\) there is $\rho$ resonance. Theory agrees with data well.
For the channel of s-wave and \(I=0\) a $0^{++}$ state is needed. This state could
be introduced to the theory.
 
For $\pi-K$ scattering the $K^*$ resonance plays a role. The comparison between theory
and data is shown in Ref.[4].
 
\section{\bf Axial-vector current}
The axial-vector part of the weak interactions between W bosn($A^i_\mu$) and mesons
is derived as
\[{\cal L}^{A}=
-{g_{W}\over 4}cos\theta_{C}
{1\over f_{a}}\{-{1\over 2}(\partial_{\mu}A^{i}_{\nu}
-\partial_{\nu}A^{i}_{\mu})(\partial_{\mu}a^{i}_
{\nu}-\partial_{\nu}a^{i}_{\mu})+A^{i\mu}j^{iW}_{\mu}\}\]
\[-{g_{W}\over 4}cos\theta_{C}
\Delta m^{2}f_{a}A^{i}_{\mu}a^{i\mu}
-{g_{W}\over4}cos\theta_{C}
f_{\pi}A^{i}_{\mu}\partial^{\mu}\pi^{i},\]
where
\[f_{a}=g^{-1}(1-{1\over2\pi^{2}g^{2}})^{-{1\over2}},\;\;\;
\Delta m^{2}=6m^{2}g^{2}.\]
Because of the axial-vector symmetry breaking there are two more terms than
VMD in the axial-vector current. Three terms together makes PCAC satisfied.
\section{weak decays}
In weak interaction of mesons there are both vector and axial-vector currents.
In this theory both the vector and the axial-vector currents are bosonized.
CVC and PCAC are satisfied.
 
In $\tau$ mesonic decays besides vector and axial-vector currents many meson
vertices are involved. All these three parts are fixed in this theory.
Systematic studies of $\tau$ mesonic decays have been done without any
adjustable parameter[5]. Theory agrees well with data.
 
Kaon weak decays, $K_{l2}$(determining $F_K$)[6], $K_{l3}$[1], $K_{l4}$[7],
and $K\rightarrow e\nu\gamma$[5] have been studied and theory agrees with data.
 
\(\Delta I={1\over2}\) rule in $K\rightarrow 2\pi$ stands there for almost
half century. Theoretical understanding is still lacking. $K\rightarrow2\pi$
are very complicated processes. At order of $O(N_C)$ there are tree diagram,
one loop, and two loop diagrams. All these diagrams are
calculable. The good news is that there are diagrams which contribute to
\(I= 1\) amplitude only.
 
\section{\bf Anomaly}
The Effective Lagrangian ${\cal L}_{IM}$ is the anomalous action of the
mesons, which contains odd number of $\gamma_5$.
 
Vertex $\pi\omega\rho$ is anomalous. It is obtained from
\begin{equation}
{1\over g}\omega^\mu<\bar{\psi}\gamma_{\mu}\psi>=\frac{-i}{(2\pi)^{D}}
\int d^{D}p
Tr\gamma_{\mu}s_{F}(x,p)\omega^\mu.
\end{equation}
The anomalous part of this vertex is obtained
\begin{eqnarray}
\lefteqn{{1\over g}\omega^{\mu}<\bar{\psi}\gamma_{\mu}\psi>=
\omega^{\mu}\partial^{2}\omega_{\mu}+
\frac{N_{c}}{(4\pi)^{2}g}{2\over 3}\varepsilon^{\mu\nu\alpha\beta}
\omega_{\mu}Tr\partial_{\nu}UU^{\dag}\partial_{\alpha}UU^{\dag}
\partial_{\beta}UU^{\dag}}\nonumber \\
 & &+\frac{N_{c}}{(4\pi)^{2}}{2\over g}\varepsilon^{\mu\nu\alpha\beta}
\partial_{\mu}\omega_{\nu}Tr\{{i\over g}[\partial_{\beta}UU^{\dag}
(\rho_{\alpha}+a_{\alpha})-\partial_{\beta}U^{\dag}U(\rho_{\alpha}
-a_{\alpha})]\nonumber \\
& &-{2\over g^{2}}(\rho_{\alpha}+a_{\alpha})U(\rho_{\beta}-a_{\beta})
U^{\dag}-{2\over g^{2}}\rho_{\alpha}a_{\beta}\}.
\end{eqnarray}
This formula is exactly the same as the one obtained by the Syracuse
group and by Witten.
${\cal L}_{\omega\rho\pi}$ is derived
\begin{equation}
{\cal L}_{\omega\rho\pi}=-\frac{N_{c}}{\pi^{2}g^{2}f_{\pi}}
\varepsilon^{\mu\nu\alpha\beta}\partial_{\mu}\omega_{\nu}
\rho^{i}_{\alpha}\partial_{\beta}\pi^{i},
\end{equation}
\[\Gamma(\omega\rightarrow 3\pi)=7.7MeV,\;\;exp.=7.49(1\pm0.02)MeV.\]
 
\[\Gamma(\omega\rightarrow\gamma\pi)=724keV.\]
The experimental value is $717(1\pm 0.07)$keV.
\[\Gamma(\rho\rightarrow\gamma\pi)=76.2keV.\]
The experimental data is $68.2(1\pm 0.12)keV$.
 
Using VMD,
\begin{equation}
{\cal L}_{\pi^{0}\rightarrow\gamma\gamma}=-{\alpha\over \pi
f_{\pi}}\varepsilon^{\mu\nu\alpha\beta}\pi^{0}\partial_{\mu}
A_{\nu}\partial_{\alpha}A_{\beta},
\end{equation}
This is ABJ anomaly.
 
The form factor of $\pi^0\rightarrow\gamma\gamma^*$ is obtained[8]
\begin{eqnarray}
F_\pi(q^2) & = & f_{\pi\rho\omega}(q^2)
{1\over2}\{\frac{-m^2_\rho+i\sqrt{q^2}\Gamma_\rho(q^2)}{q^2-m^2_\rho+i\sqrt{
q^2}
\Gamma_\rho(q^2)}+
\frac{-m^2_\omega+i\sqrt{q^2}\Gamma_\omega(q^2)}{q^2-m^2_\omega+i\sqrt{q^2}
\Gamma_\omega(q^2)}\},\nonumber \\
&&f_{\pi\rho\omega}(q^2)=
1+\frac{g^2}{2f^2_\pi}(1-{2c\over g})^2q^2.
\end{eqnarray}
$f_{\pi\rho\omega}$ is the intrinsic form factor of $\pi^0\gamma\gamma^*$.
For very low momentum we obtain
\begin{eqnarray}
F_\pi(q^2) & = & 1+a{q^2\over m^2_\pi},\\
&&a=
{m^2_\pi\over2}({1\over m^2_\rho}+{1\over m^2_\omega})
+{m^2_\pi\over2f^2_\pi}g^2(1-{2c\over g})^2.
\end{eqnarray}
The first term comes from the $\rho$ and $\omega$ poles
and the second term comes
from the intrinsic form factor of $\pi^0\gamma\gamma^*$, which is from
anomaly too.
\[a=0.0303+0.0157=0.046.\]
 
There are other anomalous processes for $\eta$ and $\eta'$. Theory agrees
with the data.
 
$K^*\rightarrow K\pi\pi$ decay originates in anomaly and is calculated
\[\Gamma=2.65keV\]
which is below the experimental limit.
\section{\bf ChPT}
The ChPT is constructed by chiral symmetry. This theory is working when
$E<500MeV$.
To the fourth order in
derivatives there are 10 parameters which are determined by fitting data.
Any effective theory for higher energy should take the ChPT as a low
energy limit, namely the 10 parameters should be predicted. Indeed,
the effective large $N_C$ QCD of mesons goes back to ChPT(E<$m_rho$)
and the 10 parameters are predicted[6]. \\
We predicted all the 10
coefficients by using $\pi\pi$ and $\pi K$ scatterings, masses of
pseudoscalar mesons, radius of pion, form factors of $\pi\rightarrow
e\gamma\nu$.
Up to the second order in current quark masses, the expressions of
pseudoscalar mesons are the same as obtained by ChPT.
\begin{table}[h]
\begin{center}
\caption{Predictions of the Values of the coefficients}
\begin{tabular}{|c|c|c|c|c|c|c|c|c|c|}  \hline
$10^{3}L_{1}$&$10^{3}L_{2}$&$10^{3}L_{3}$
&$10^{3}L_{4}$&$10^{3}L_{5}$&$10^{3}L_{6}$
&$10^{3}L_{7}$&$10^{3}L_{8}$&$10^{3}L_{9}$
&$10^{3}L_{10}$  \\ \hline
1.0&2.0&-5.16&0&4.77&0&0&-0.079&8.3&-7.1\\ \hline
\end{tabular}
\end{center}
\end{table}
 
%\begin{table}[h]
%\begin{center}
%\caption{Values of the coefficients}
%\begin{tabular}{|l|l|l|l|l|l|l|l|l|l|} \hline
%$10^{3}L_{1}$&$10^{3}L_{2}$&$-10^{3}L_{3}$
%&$-10^{3}L_{4}$&$10^{3}L_{5}$&-$10^{3}L_{6}$
%&-$10^{3}L_{7}$&$10^{3}L_{8}$&$10^{3}L_{9}$
%&-$10^{3}L_{10}$  \\ \hline
%$.4\pm.3$&$1.4\pm.3$&$3.5\pm1.1$&$.3\pm.5$&
%$1.4\pm.5$&$.2\pm.3$&$.4\pm.2$&$.9\pm.3$&
%$6.9\pm.7$&$5.5\pm.7$ \\ \hline
%$.52\pm.23$&$.7\pm.2$&$2.7\pm.99$&&
%$.65\pm.12$&&$.26\pm.15$&$.47\pm.18$&&
% \\ \hline
%\end{tabular}
%\end{center}
%\end{table}
%The first column is the fit of ChPT and the second is the fit by Pich et al., in
% which $6^{th}$ order derivatives are taken into account.
 
In this theory the quark condensate(12)
is obtained
\begin{equation}
<\bar{u}u>=<\bar{d}d>=<\bar{s}s>=-(0.378GeV)^3.
\end{equation}
A larger quark condensate leads to greater value
of $L_5$.
On the other hand, \({f_K\over f_\pi}=1.19\)
is obtained in this paper, which is $2.5\%$ away from the data.
Because of the same reason a smaller $L_8$ is obtained in this paper.
 
Small current quark masses are obtained
\[m_u=0.91MeV,\;\;m_d=2.15MeV,\;\;m_s=52.3MeV.\]
As a check, the effect of current quark mass in $\pi^0\rightarrow
\gamma\gamma$ is calculated
\begin{eqnarray}
{\cal L}_{\pi^0\rightarrow\gamma\gamma} & = & -\frac{\alpha}
{\pi f_\pi}f_q
\pi^0\varepsilon^{\mu\nu\lambda\beta}
\partial_\mu A_\nu\partial_\lambda A_\beta,\\
&&f_q= 1-{1\over m}(m_u+m_d)+
{g^2\over2f^2_\pi}(1-{2c\over g})^2 m^2_{\pi^0},\\
&&\Gamma(\pi^0\rightarrow\gamma\gamma)=\frac{\alpha^2 m^3_\pi}{16\pi^3
f^2_\pi}
\{1-{1\over m}(m_u+m_d)+
{g^2\over2f^2_\pi}(1-{2c\over g})^2 m^2_{\pi^0}\}^2
\end{eqnarray}
\begin{equation}
f_q
=1+4.97\times10^{-3}.
\end{equation}
The decay width of
$\pi^0\rightarrow\gamma\gamma$ is increased by $1\%$.
\begin{equation}
\Gamma_{\pi^0\rightarrow\gamma\gamma}=7.89eV.
\end{equation}
The data is $7.83(1\pm0.071)eV$.
\section{\bf OZI rule}
At the tree level $\phi\rightarrow\rho\pi$ if forbidden. Only loop
diagrams of mesons contribute to this decay. At tree level, the decay
amplitude is $O(N_C)$ and at the loop level of this decay the order is
$O(1)$. It is well known that the decay $\phi\rightarrow\rho\pi(3\pi)$
is forbidden
by OZI rule. In this theory there is no tree diagram for this decay mode and
only loop diagrams of mesons contribute to this decay mode. The order of the
amplitude of this decay mode is $O(1)$.
Because the cutoff determined is about 1.8GeV
,therefore, only low lying meson states contribute to the loop diagrams
of the decay, $\phi\rightarrow\rho\pi(3\pi)$.
This decay is calculable. It is
a serious test on this effective chiral large $N_C$ theory.
\section{Summary}
This effective theory is phenomenologically successful. It has most nice
features
obtained in previous studies. Especially, this theory uses $N_C$ expansion
to do concrete physical study. On the other hand, there are more work ahead:
\begin{enumerate}
\item $K\rightarrow2\pi$ decays and \(\Delta I={1\over2}\) rule,
\item CP violation in rare kaon decays,
\item OZI rule suppressed decays $\phi\rightarrow\rho\pi(3\pi)$,
\item Loop corrections, such as in $\rho\rightarrow2\pi$...et al.,
\item Current quark mass corrections,
\item The match between this theory and perturbative QCD.
\end{enumerate}
{\bf Acknowledgement}\\
This research was supported by DOE Grant No. DE-FG02-00ER45846.

\pagebreak
\begin{flushleft}
{\bf Figure Captions}
\end{flushleft}
{\bf FIG. 1.} Pion form factor in time-like region.
\\ {\bf FIG. 2.} Pion form factor in space-like region.
\\{\bf FIG. 3.} Pion form factor in space-like region.
\\{\bf FIG. 4.} Charged kaon form factor in time-like region.
\\{\bf FIG. 5.} Charged kaon form factor in space-like region.
\\{\bf FIG. 6.} Phase shift $\delta^{{1\over2}}_1$.
\\{\bf FIG. 7.} Phase shift $\delta^{{1\over2}}_1$.
\\{\bf FIG. 8.} Phase shift $\delta^{{3\over2}}_0$.
\\{\bf FIG. 9.} $K^+\pi^-$ p-wave cross section.

\begin{figure}
\begin{center}
\includegraphics[width=7in, height=7in]{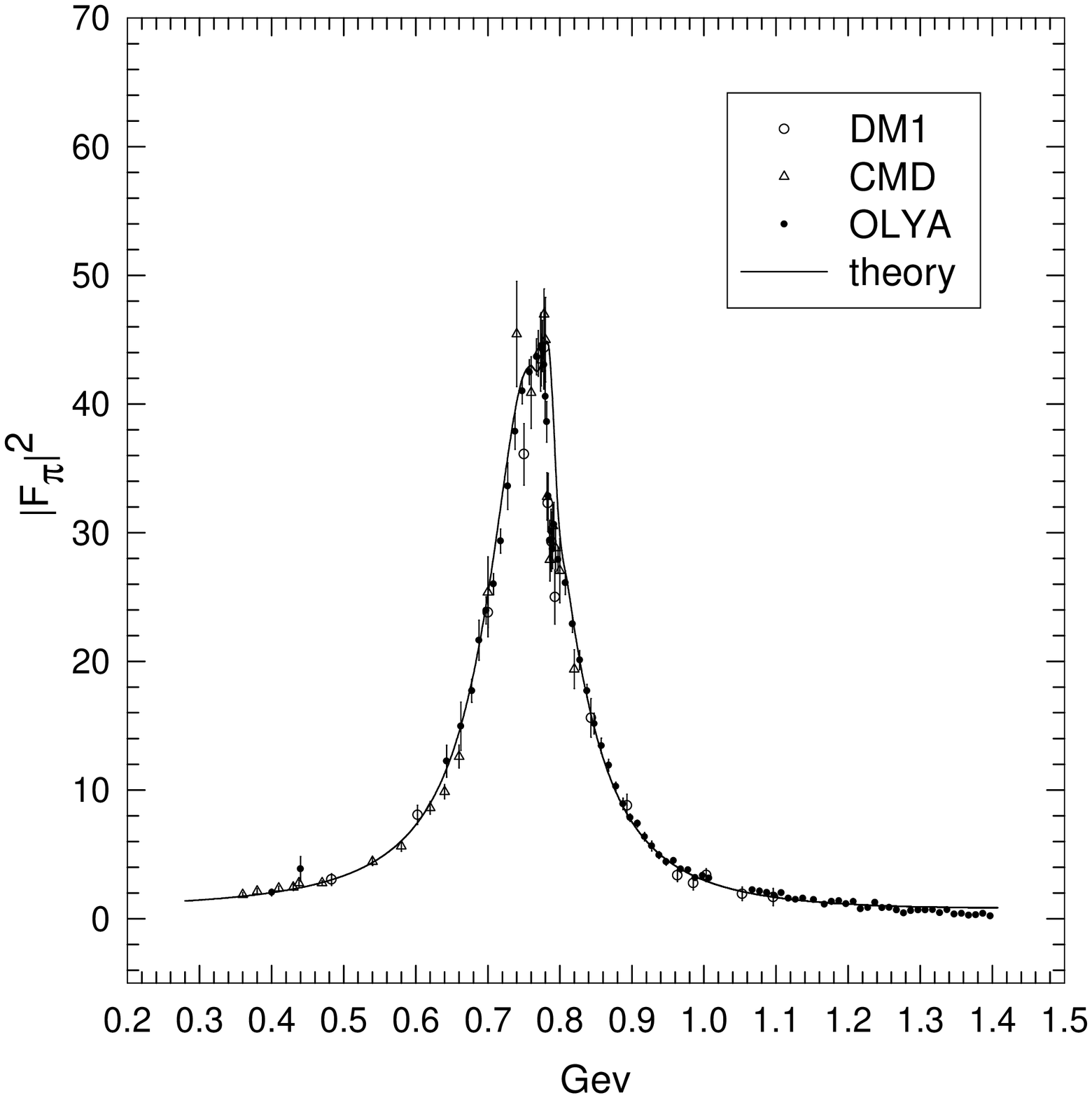}
FIG. 1.
\end{center}
\end{figure}

\begin{figure}
\begin{center}
\includegraphics[width=7in, height=7in]{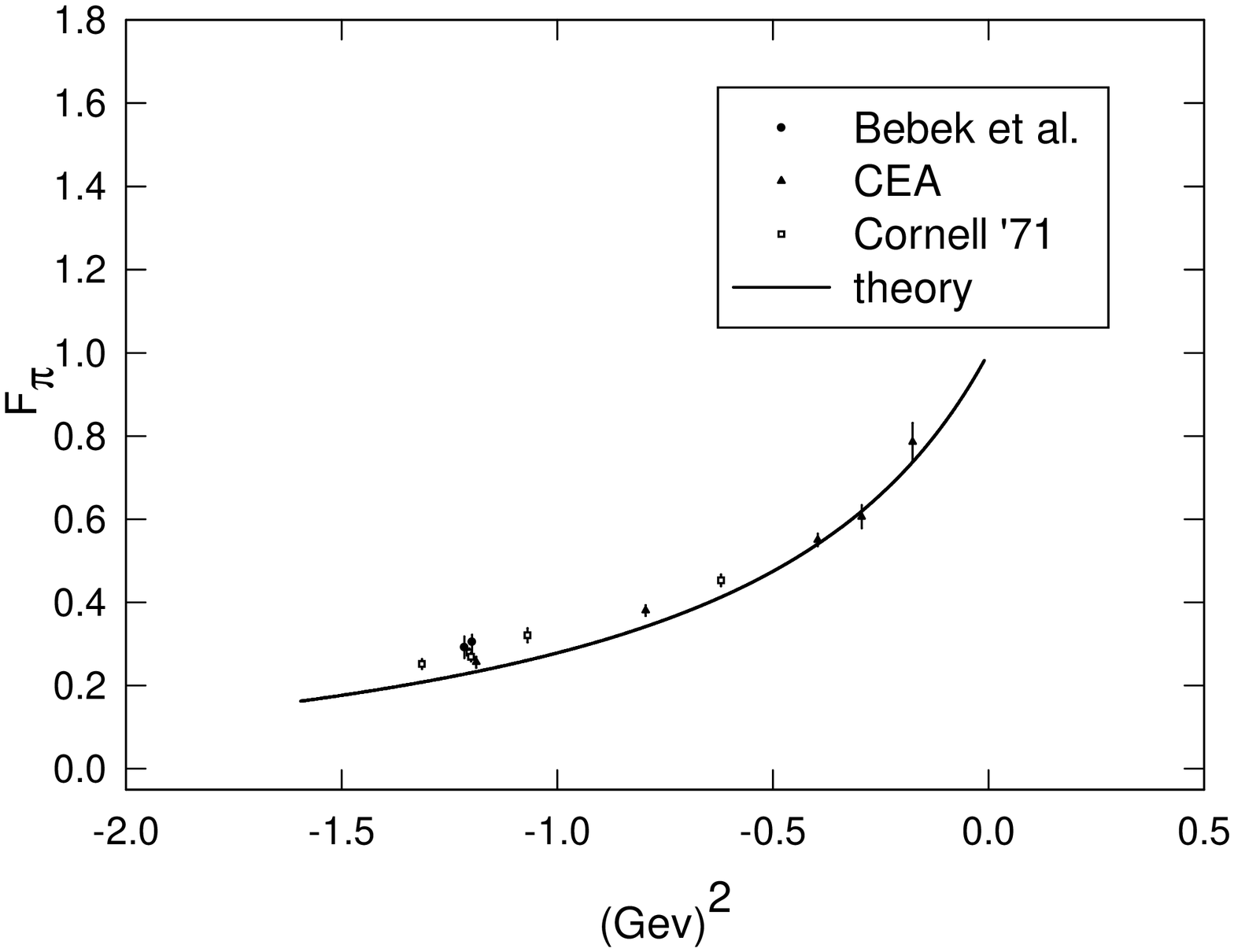}
FIG. 2.
\end{center}
\end{figure}

\begin{figure}
\begin{center}
\includegraphics[width=7in, height=7in]{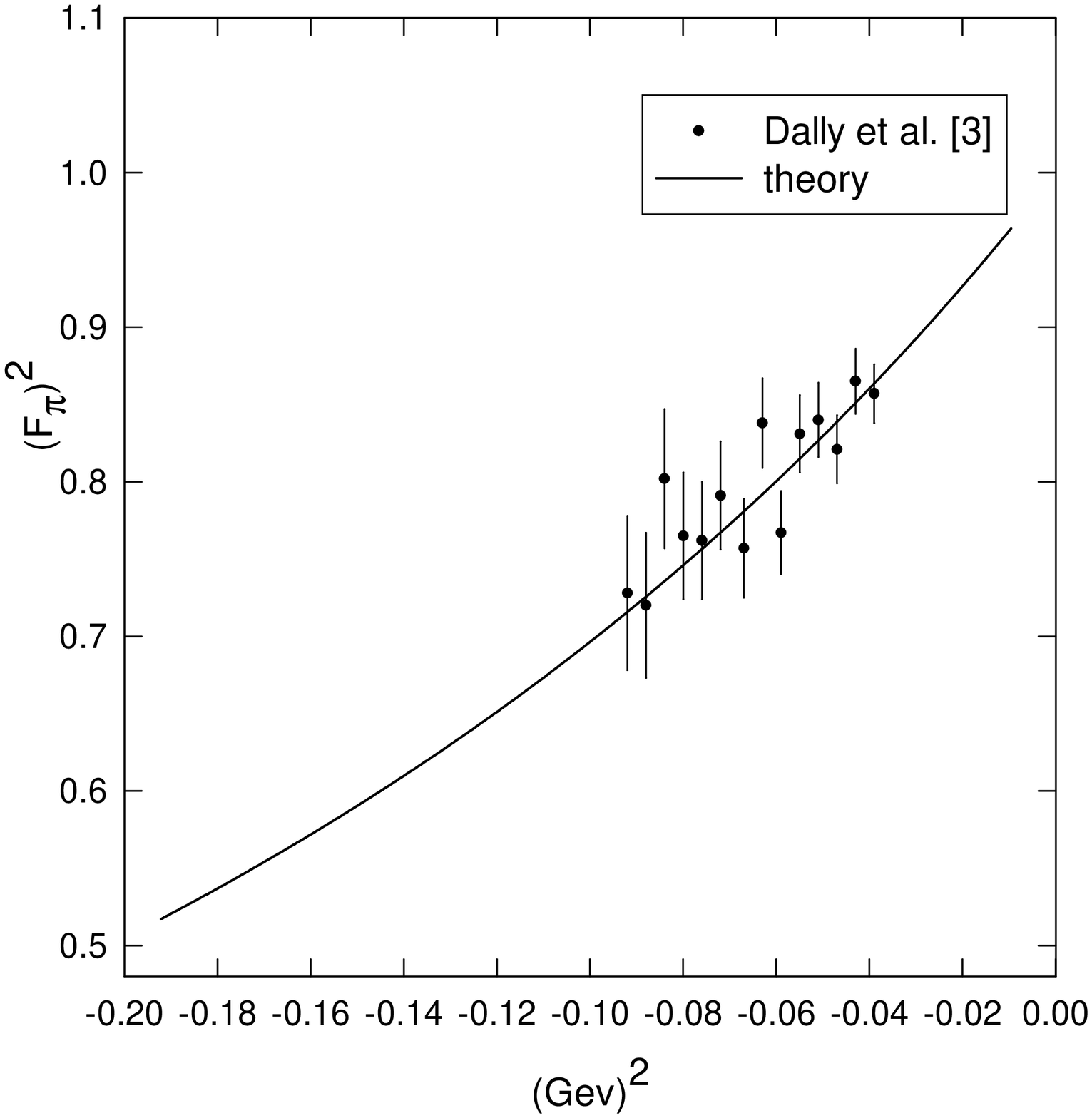}
FIG. 3.
\end{center}
\end{figure}

\begin{figure}
\begin{center}
\includegraphics[width=7in, height=7in]{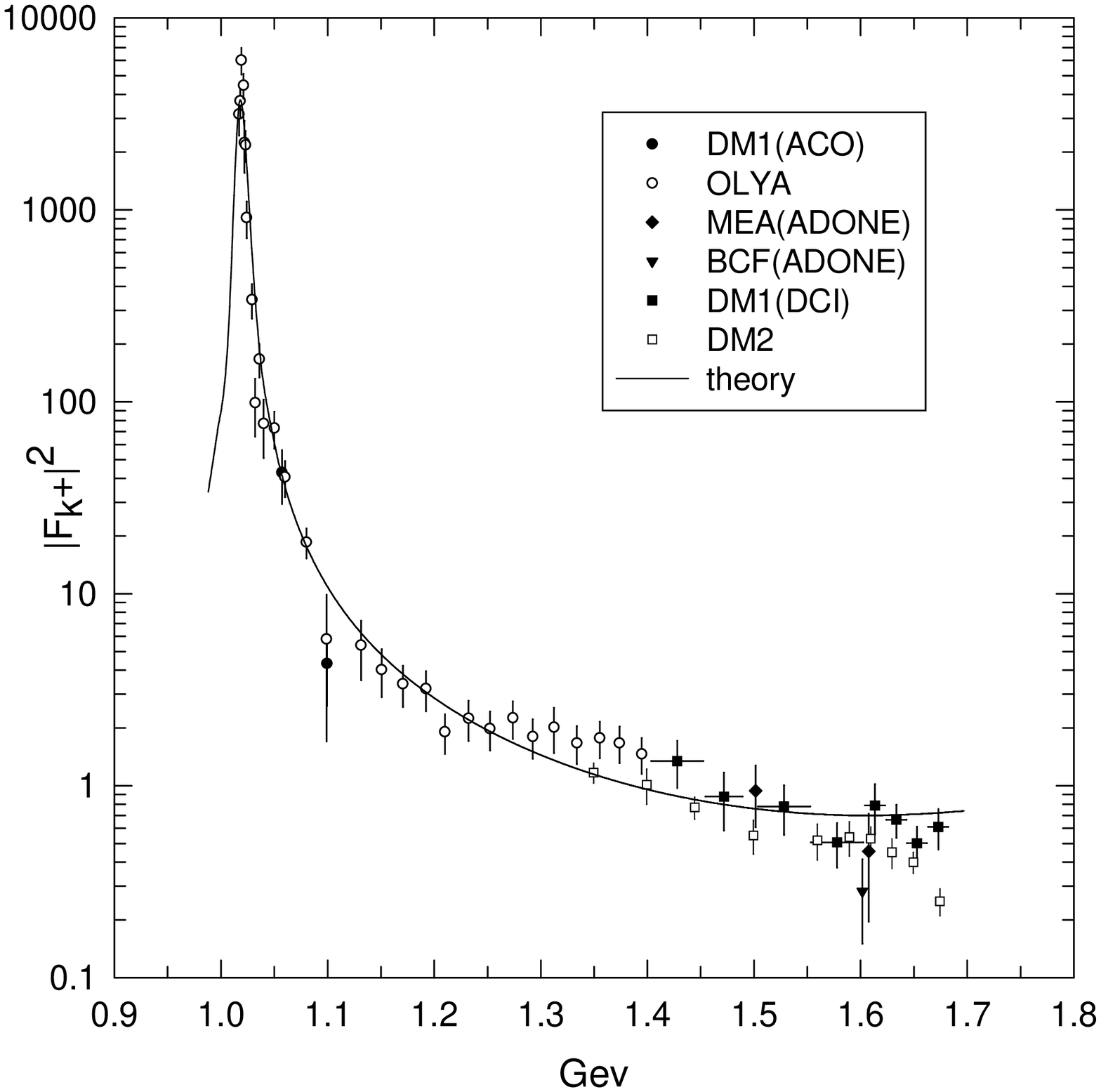}
FIG. 4.
\end{center}
\end{figure}

\begin{figure}
\begin{center}
\includegraphics[width=7in, height=7in]{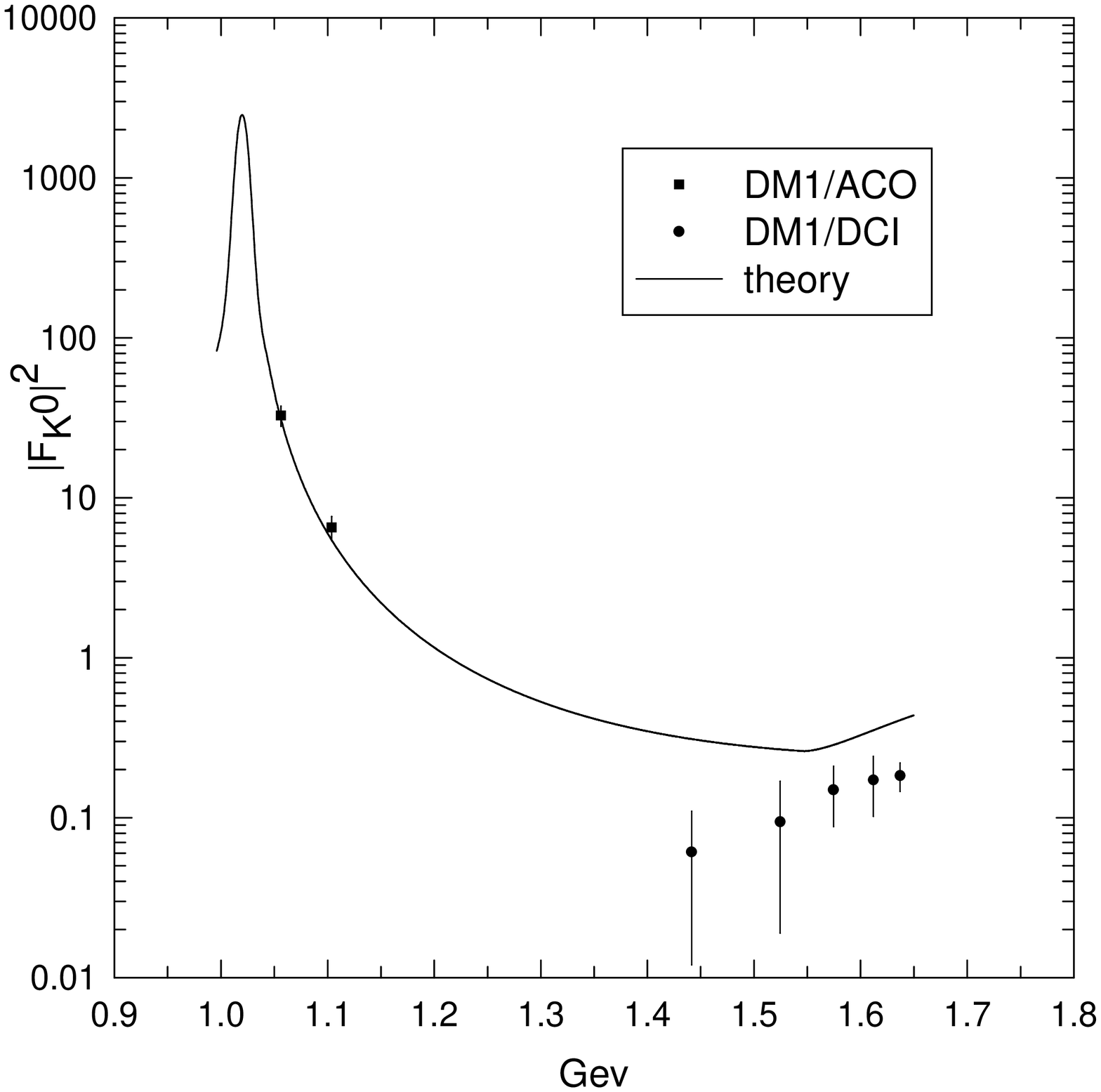}
FIG. 5.
\end{center}
\end{figure}

\begin{figure}
\begin{center}
\includegraphics[width=7in, height=7in]{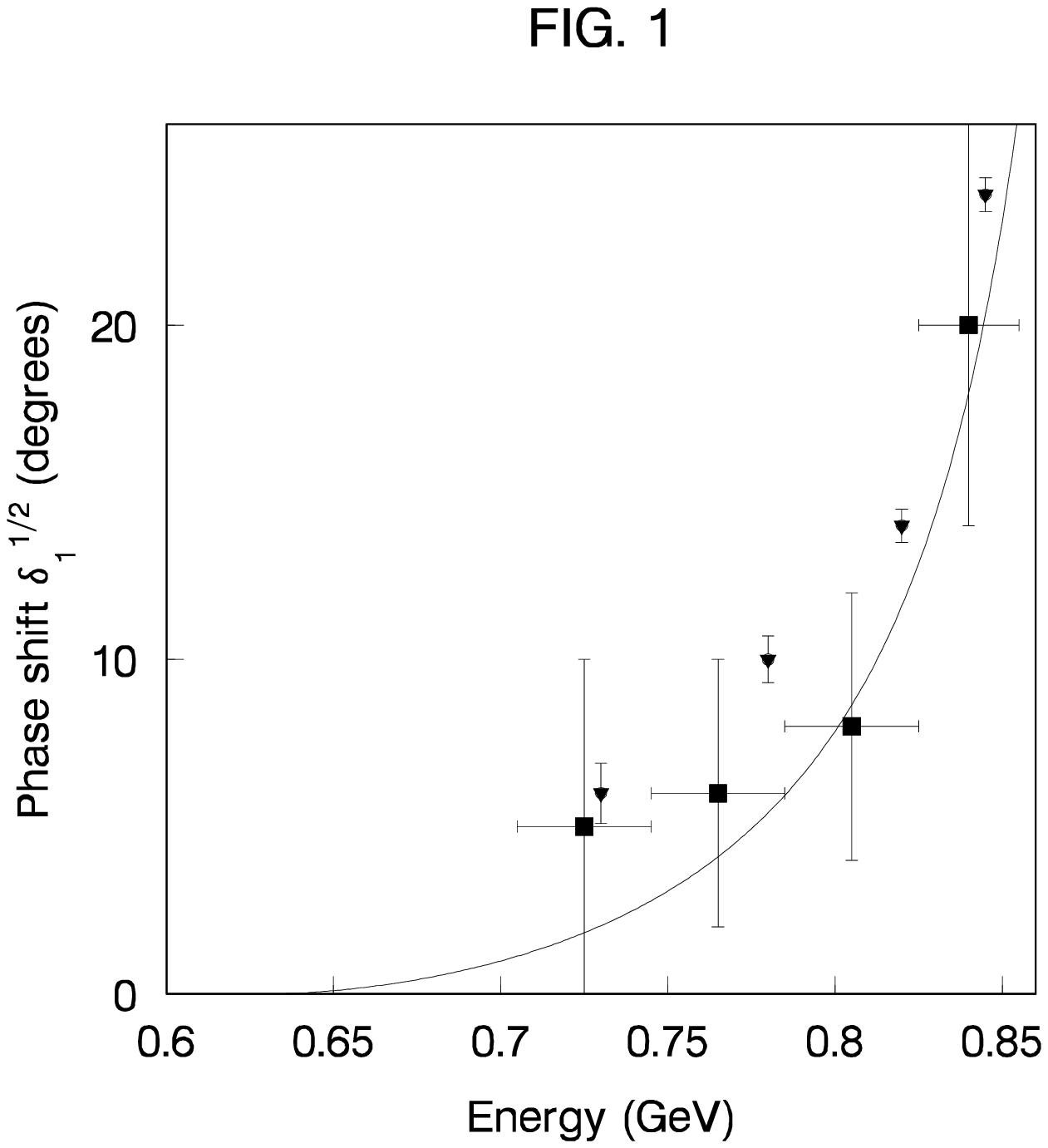}
FIG. 6.
\end{center}
\end{figure}
\begin{figure}
\begin{center}
\includegraphics[width=7in, height=7in]{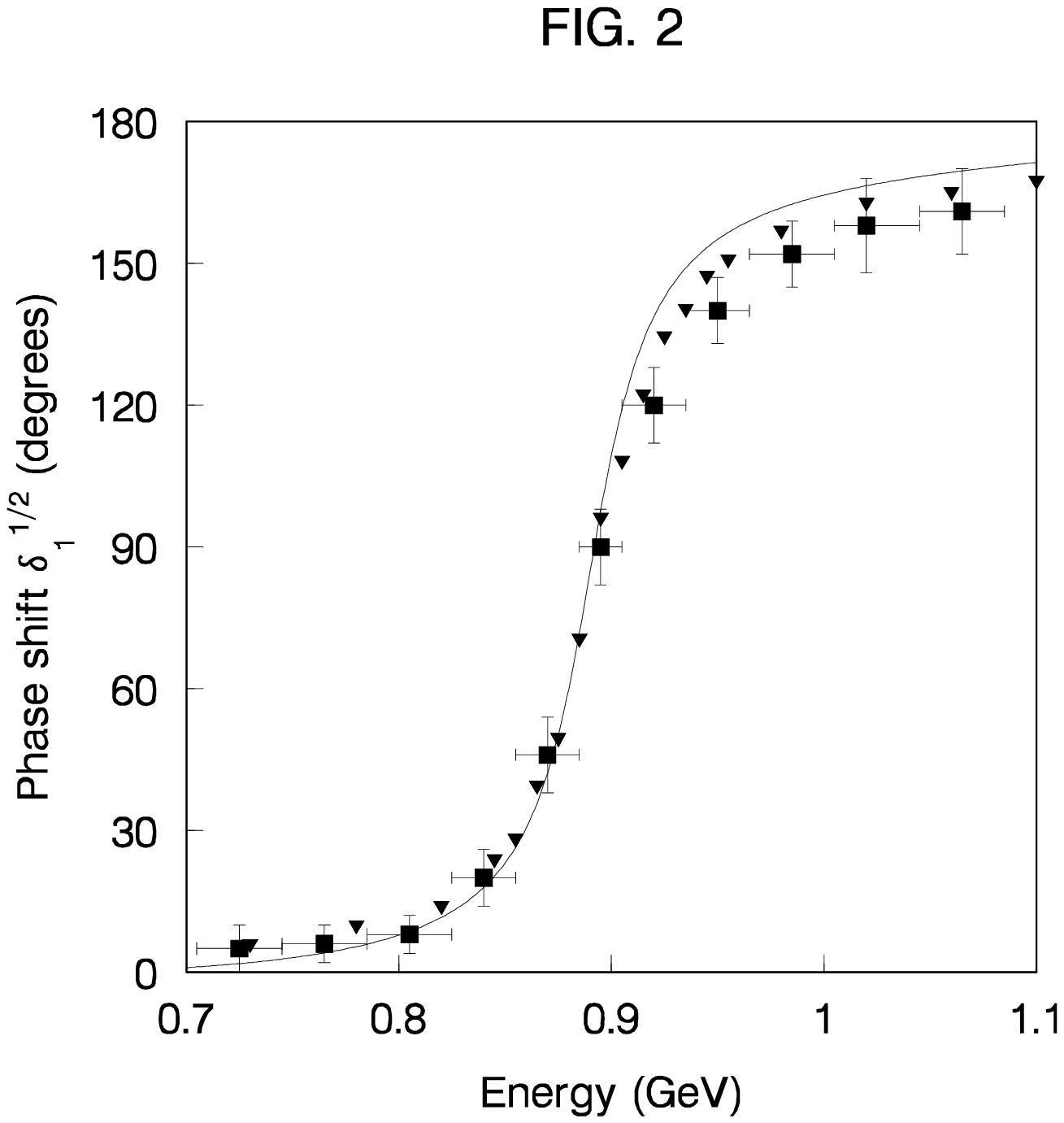}
FIG. 7.
\end{center}
\end{figure}
\begin{figure}
\begin{center}
\includegraphics[width=7in, height=7in]{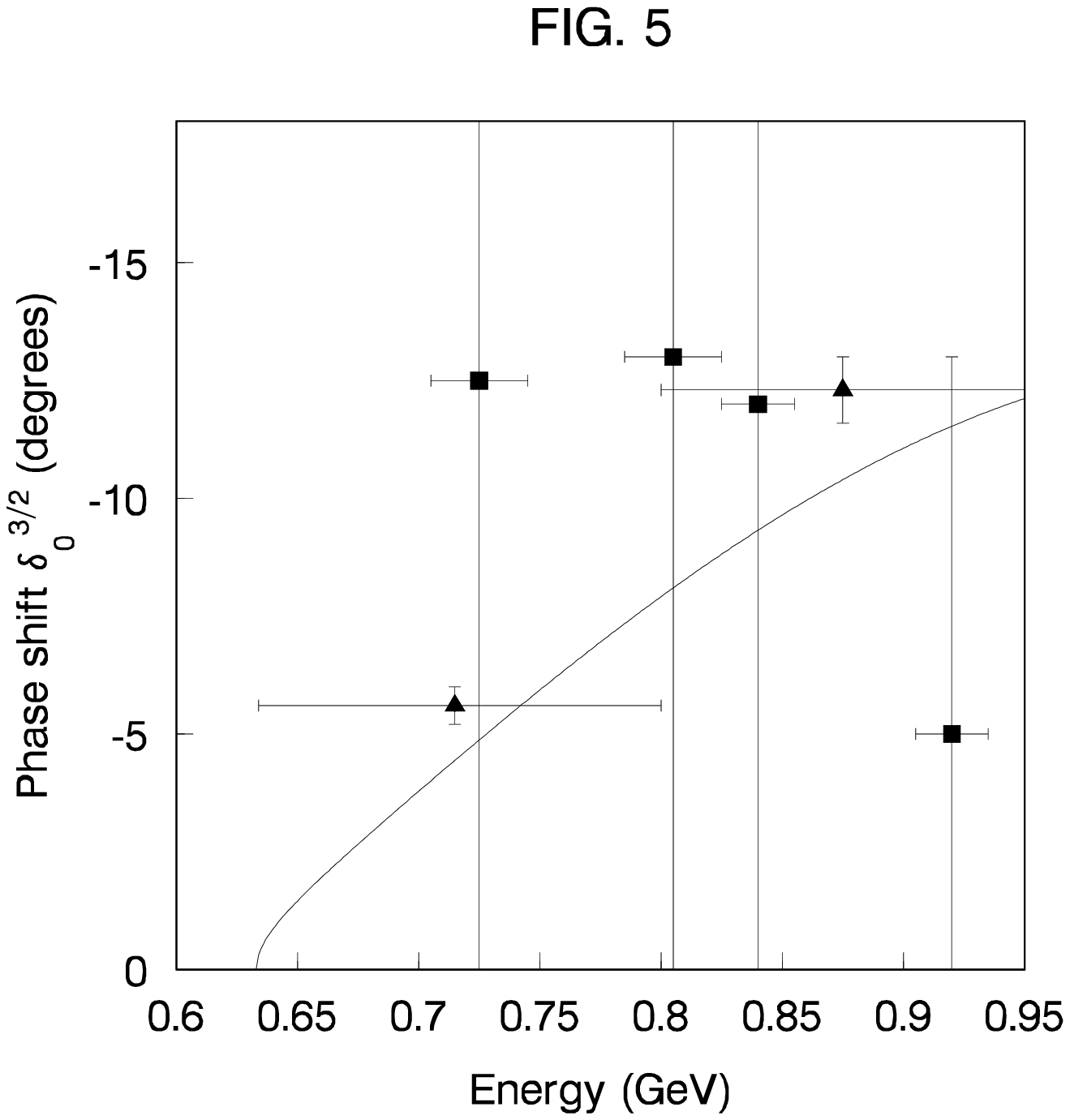}
FIG. 8.
\end{center}
\end{figure}
\begin{figure}
\begin{center}
\includegraphics[width=7in, height=7in]{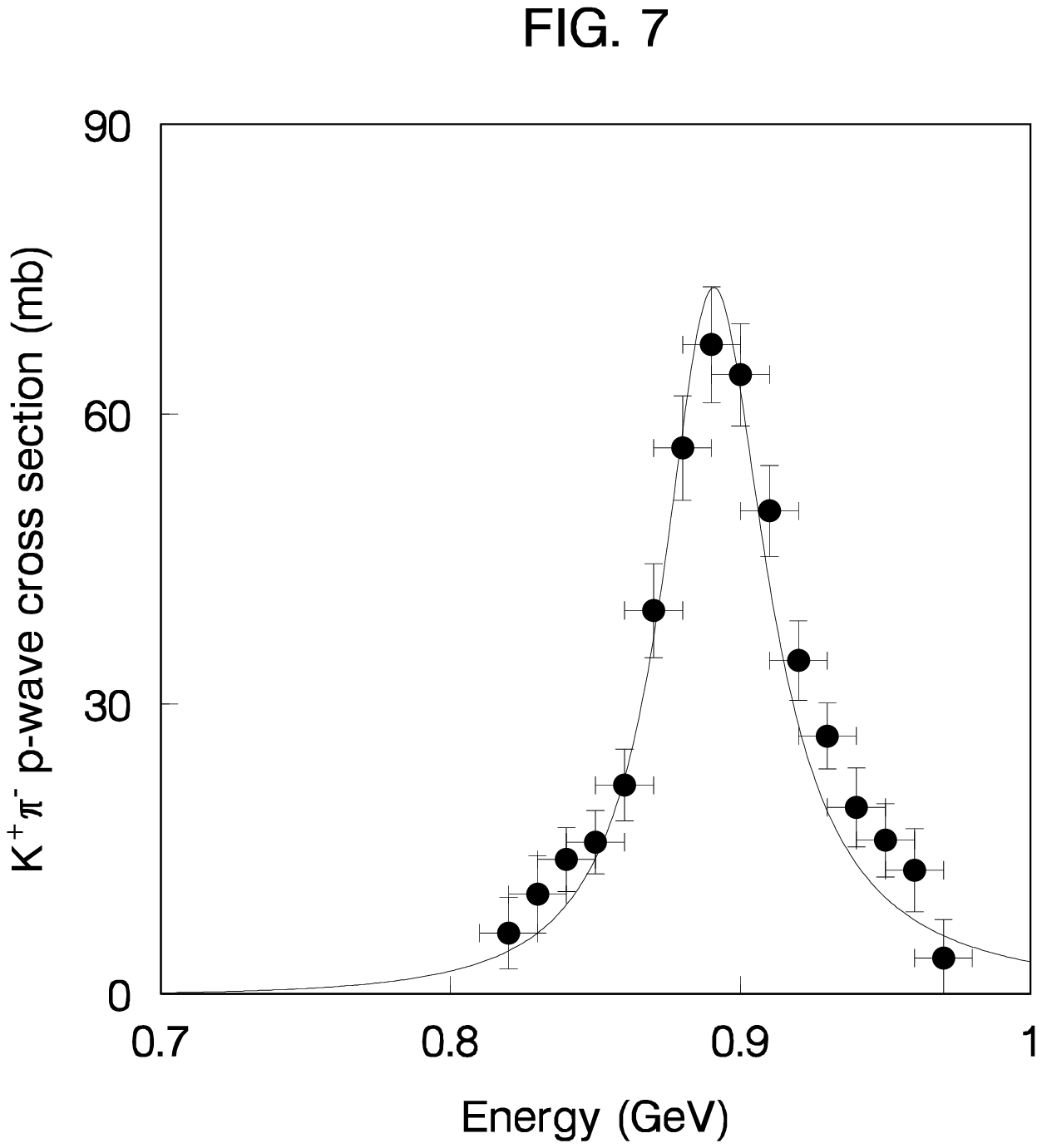}
FIG. 9.
\end{center}
\end{figure}

\end{document}